\documentclass[letterpaper]{article}

\usepackage[T1]{fontenc}

\usepackage{geometry}
\usepackage{setspace}

\usepackage[style = chem-acs]{biblatex}
\usepackage{graphicx}
\usepackage{float}
\newfloat{scheme}{htbp}{los}
\floatname{scheme}{Scheme}
\floatname{chart}{Chart}
\newfloat{graph}{htbp}{loh}

\usepackage{chemformula} 
\usepackage[version = 4]{mhchem} 

\usepackage{comment}
\usepackage{amssymb}
\usepackage{amsfonts}
\usepackage{amsmath}
\usepackage{amsxtra}
\usepackage{amscd}
\usepackage{bm}
\usepackage{amsthm}
\usepackage{graphicx}
\usepackage{epsf}
\usepackage{bbold}
\usepackage{xcolor}
\usepackage{mathalfa}
\usepackage{makecell} 
\usepackage{multirow}
\usepackage{hhline}
\usepackage{tikz}
\usepackage{physics}
\usepackage{tabularx}
\usepackage{eucal}
\usepackage{stackengine}

\newcommand{\beq}{\begin{equation}}
\newcommand{\eeq}{\end{equation}}
\newcommand{\bmul}{\begin{multline}}
\newcommand{\emul}{{\end{multline}}}
\newcommand\beqa{\begin{eqnarray}}
\newcommand\eeqa{\end{eqnarray}}
\newcommand\bea{\begin{array}}
\newcommand\eea{\end{array}}
\newcommand\ba{\begin{array}}
\newcommand\ea{\end{array}}

\newcommand{\neqa}{\nonumber\end{eqnarray}}

\usepackage[export]{adjustbox}

\makeatletter
\protected\def\xvcenter{%
  \hbox\bgroup$\everyvbox{\everyvbox{}\aftergroup\m@th\aftergroup$\aftergroup\egroup}%
  \vcenter
}

\DeclareRobustCommand{\midscript}[1]{
  \mathchoice{\mid@script\scriptstyle{#1}}
    {\mid@script\scriptstyle{#1}}
    {\mid@script\scriptscriptstyle{#1}}
    {\mid@script\scriptscriptstyle{#1}}
}
\newcommand{\mid@script}[2]{
  \vcenter{\hbox{$\m@th#1#2$}}
}

\usepackage{authblk}

\title{Resonant Structure of Second Harmonic Generation in Multilayer Graphene Polytypes}

\author[1,2]{Patrick Johansen Sarsfield$^*$}
\affil[1]{National Graphene Institute, University of Manchester, Manchester M13 9PL, United Kingdom}
\affil[2]{Department of Physics \& Astronomy, University of Manchester, Manchester M13 9PL, United Kingdom}

\author[3]{Takaaki V. Joya}
\affil[3]{Department of Physics, University of Osaka, Toyonaka, Osaka 560-0043, Japan}
\author[3]{Takuto Kawakami}
\author[3]{Mikito Koshino}

\author[1,2]{Vladimir Fal'ko}

\date{*patrick.sarsfield@manchester.ac.uk}

\begin{document}

\maketitle

\begin{abstract}
Second harmonic generation (SHG) is a powerful optical tool for identifying non-centrosymmetric crystalline structures. Here, we analyze SHG in multilayer graphenes (MLG), with a focus on its dependence on the stacking order, encapsulation environment and biasing which break inversion symmetry in multilayers, as well as the SHG sensitivity to the electron-hole asymmetry in the MLG spectra and doping. In particular, we identify stacking-order-dependent resonant features in the SHG spectra for trilayers and tetralayers, suggesting that infra-red range SHG offers a non-invasive characterization method for distinguishing between MLG polytypes, as well as optical identification of crystallographic direction in MLG films.

\end{abstract}

\maketitle

\section{Introduction}

Non-linear optical responses have been widely used as a probe to study the symmetry, topology, and quantum geometry~\cite{NagaosaMorimoto2017,OrensteinMooreMorimoto2021,MaSong2023,PengXiuWang2025} in various two-dimensional (2D) materials~\cite{LiHeinz2013,YangSongYao2020}.
In particular, second harmonic generation (SHG), which is a second-order optical response where the input frequency is doubled in the output~\cite{FrankenWeinreich1961,Kleinman1962,SipeShkrebtii2000}, has been actively employed as a non-contact probe to identify features such as strain~\cite{LiangLiuYu2017,MennelPaurMueller2019,LuSunNam2023}, stacking order~\cite{KimMullerPark2013,ShanLieWu2018,YaoBasovSchuck2022,ZhouZhuXu2024,roy2025detecting},  domains~\cite{HsuChouChang2014,KhanRahmanLu2022,YaoBasovSchuck2022,AokiIdeueIwasa2024,Yao2021} and defects~\cite{CarvalhoMalard2019,CunhaAharanovichMalard2020} in 2D crystals and moir\'{e} materials.
The above were studied in a variety of 2D materials, including multilayer graphene~\cite{DeanDriel2009,DeanDriel2010,SongTongZhang2024} (MLG), hexagonal boron nitride~\cite{Yao2021,KimMullerPark2013,CunhaAharanovichMalard2020}, transition metal dichalcogenides~\cite{HsuChouChang2014,LiangLiuYu2017,MennelPaurMueller2019}, and ferroelectric materials~\cite{KhanRahmanLu2022,AbdelwahabMaierLoh2022,AokiIdeueIwasa2024,vizner2025sliding}.

In the context of graphene studies, experimental efforts have already been made to use SHG to identify the stacking order of trilayers~\cite{ShanLieWu2018}, tetralayers~\cite{ZhouZhuXu2024}, and even pentalayers ~\cite{roy2025detecting}.
Typically, those experiments detected an SHG signal at a single input frequency of light, aiming to distinguish the polytypes from other SHG-inactive stackings that have inversion symmetry. Here we propose a comprehensive SHG theory incorporating symmetry breaking from non-centrosymmetric stacking and environmental encapsulation, electron-hole asymmetry, and carrier doping. For this we use the hybrid $\vb{k}\cdot\vb{p}$ theory -- tight-binding model~\cite{McClure_Band_1957,Slonczewsi_Band_1958}, fully parameterized for tri-, and tetralayer graphenes~\cite{mcellistrim_spectroscopic_2023,sarsfield2024substrate,koshino2009gate}, combined with the self consistent evaluation of electrostatic charge distribution across layers based on the Hartree approximation~\cite{Slizovskiy_dielectric_2021}. The influence of the encapsulation environment is accounted for by the on-layer proximity potentials ~\cite{sarsfield2024substrate,boschi_built-bernal_2024}.

\section{Methods}

\subsection{SHG phenomenology and selection rules set by MLG lattice symmetry}

\begin{figure*}[]
\centering
  { \includegraphics[width=\textwidth]{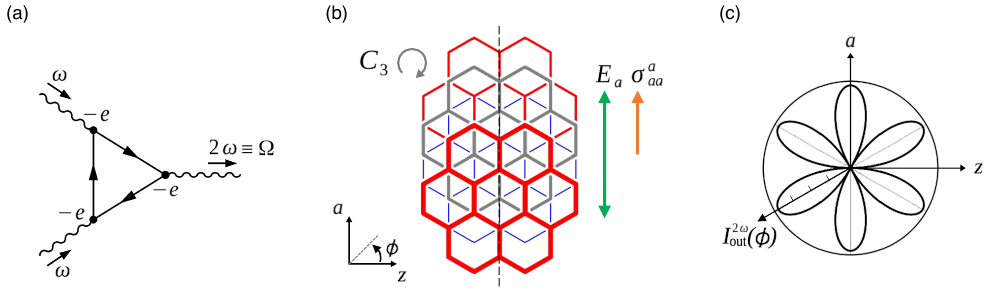}}
\caption{\label{fig:diagramlatticepolarplot} (a) The Feynman diagram for the dominant process contributing to the SHG signal in MLG. (b) Top view of ABCB tetralayer lattice, with with red, gray, and blue color marking to the `B', `C', and `A' layers, respectively (for a side view, see Fig.~\ref{fig:tetralayer_bands}), and axes identifying a pair of armchair ($a$) and zigzag ($z$) axes. This sketch shows that the crystal has a threefold in-plane rotational symmetry $C_3$ and a mirror symmetry across the vertical planes with a projection aligned with armchair directions, such as shown by gray dashed line. (c) Dependence of a $2\omega$ emission intensity, $I^{2\omega}_{\rm out}$, on the orientation of the polarization axis in a single-polarizer SHG set-up (angle $\phi$ is counted from the armchair direction).
}
\end{figure*}

In optics, SHG appears as a conversion of two $\hbar\omega$ photons into one $2\hbar\omega$ (double-frequency) photon due to the non-linear dynamical response of the material, illustrated by the Feynman diagram in Fig.~\ref{fig:diagramlatticepolarplot}(a). Phenomenologically, it can be characterized by a displacement current ($j=dP/dt$, where $P$ is polarization of the medium),
\begin{align}
    j^{2\omega}_\mu &= \sigma^\mu_{\alpha\beta}(\omega)\,E_\alpha E_\beta,
    \label{eq:photocurrent}
\end{align}
where we sum over repeated indices $\alpha,\beta,\mu=a,z$ which are used to identify projections of electric field vector, $\vb{E}$, carried by the incoming light (of frequency $\omega$) onto a pair of mutually orthogonal armchair and zigzag axes of graphene's lattice, $a$ and $z$ -- see in Fig.~\ref{fig:diagramlatticepolarplot}(b).

For a material to be SHG-active, it must have broken spatial inversion symmetry~\cite{FrankenWeinreich1961,Kleinman1962,SipeShkrebtii2000}: otherwise, $\sigma^\mu_{\alpha\beta} = 0$, as both $\vb{E}$ and $\vb{j}$ invert sign upon inversion transformation. This symmetry property of SHG makes it a signal of non-centrosymmetric crystals~\cite{ShanLieWu2018,ZhouZhuXu2024,roy2025detecting}; among multilayer graphenes, those include Bernal  trilayer and mixed-stacking tetralayer~\cite{GarciaMcEllistrimFalko2023}. In the context of 2D crystals with inversion-symmetric lattice structure (e.g., bilayer or rhombohedral few-layer graphenes) asymmetry can also be induced by an encapsulation environment, or the out-of plane electric bias. For 2D materials, where the light-matter interaction involves in-plane-polarized light, the $C_2$ symmetry (180$^\circ$ rotations) should be absent, too (for example, this forbids SHG in monolayer graphene). 

As our further analysis is focused on multilayer graphenes, which have `trigonal' $C_3$ rotational symmetry and mirror symmetry with respect to vertical planes (armchair directions in one of the monolayers), the SHG tensor in Eq.~\eqref{eq:photocurrent} should satisfy the following constraints,  
\begin{align}
    \begin{split}
        \sigma^a_{aa} &= -\sigma^z_{za} = -\sigma^z_{az} = -\sigma^a_{zz};
        \\
         \sigma^z_{zz} &= -\sigma^a_{za} = -\sigma^a_{az} = -\sigma^z_{aa} =0,
    \end{split}
    \label{eq:symmetry}
\end{align}
hence, fully characterized by one tensor component, $\sigma^a_{aa}(\omega)$. After the above relations were applied to a circularly polarized incoming light, with $\vb{E}_\pm = \frac{1}{\sqrt{2}} (\vb{l}_x\pm i \vb{l}_y) E_\pm$ (SI Section 1  for details), we find that the SHG current and, therefore, the emitted photons feature a reversed circular polarization, 
\begin{equation}
    \vb{j}^{2\omega}_\mp=\pm i\sqrt{2}\,\sigma^a_{aa}(\omega)\, (\vb{l}_x\pm i \vb{l}_y)  E^2_\pm.
\end{equation}
While such an inverted circular polarization relation between the incoming and outgoing light may look counterintuitive, it reflects angular momentum conservation rules for $C_3$-symmetric crystals~\cite{KonishiIshiiKuwataGonokami2014}, where the discrete rotational symmetry causes a Bragg-type angular momentum (that is, its out-of-plane component, $L_z$) transfer to the lattice in units of $\pm 3\hbar$ (in the same way as the crystalline periodicity leads to momentum transfer in units of reciprocal lattice vectors). Therefore, two incoming circularly polarized photons that -- together -- bring into a $C_3$-symmetric crystal the total angular momentum $\pm 2\hbar$ can generate a photon that would carry out an angular momentum of $(\pm 2 \mp 3)\hbar =\mp \hbar$, which, indeed, corresponds to the reversal of the circular polarization. 

\begin{figure*}[]
\centering
  { \includegraphics[width=0.95\textwidth]{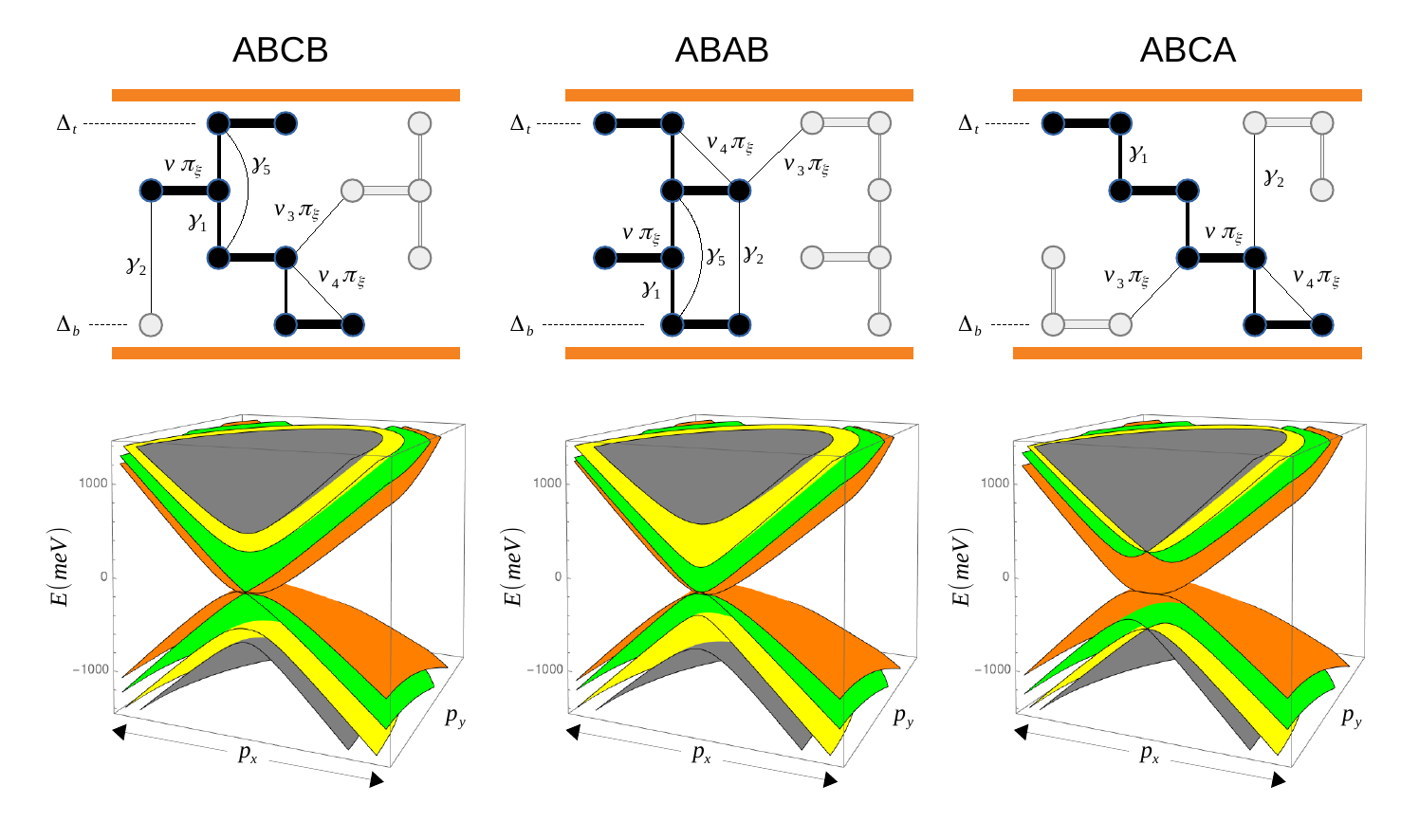}}
\caption{\label{fig:tetralayer_bands}
Top row of panels shows a side view of tetralayer polytypes lattices with all hopping terms included in the implemented MLG Hamiltonians. We use the fine dashed lines to highlight the on-layer potentials, $\Delta_{t/b}$, induced by the encapsulation environment. Stacking diagrams for ABA and ABC graphene can be obtained by removing the top or bottom layer of the ABCB tetralayer. The bottom row of panels shows the band structure of each tetralayer polytype in the energy range relevant for optical processes in the infrared frequency range (nested bands are exposed by slicing along the $p_y=0$ line).
}
\end{figure*}

In terms of the Feynman diagram in Fig.~\ref{fig:diagramlatticepolarplot}(a), excitation by circularly polarized light, $\vb{E}_+$ or $\vb{E}_-$, and the consequent circularly polarized SHG emission would invoke electronic transitions in only one of the two valleys, K$_+$ or K$_-$, of graphene's spectrum. At the same time, a more common experimental SHG setup is based on a single-polarizer measurement, where both the pumped and detected photons are linearly polarized in the same direction. In that case, the SHG process involves coherent electron transitions in both K$_+$ and K$_-$ valleys, and the polarization dependence of the SHG emission is demonstrated by the flower-like intensity diagram in Fig.~\ref{fig:diagramlatticepolarplot}(c) showing that the maximum effect,
\begin{align}
    I^{2\omega}_{\rm out}=\frac{A}{2(\varepsilon_0c)^3}\abs{\sigma^a_{aa}(\omega)}^2 (I^\omega_{\rm in})^2,
    \label{eq:shg_intensity}
\end{align}
is achieved with the polarizer aligned with one of graphene's armchair axes. Here, $I^\omega_{\rm in}$ is intensity of incoming light, $c$ is the speed of light, $\varepsilon_0$ is the permittivity of vacuum, and a numerical prefactor $A$ depends on the electromagnetic environment, e.g., due to reflection from the substrate~\cite{WangSipeCheng2024}. Details are given in SI Section 2. A similar dependence on polarization direction of the incoming light would also appear in the cross-polarized measurement configuration, due to the relation $\sigma^a_{aa} = -\sigma^a_{zz}$ between tensor components in Eq.~\eqref{eq:symmetry}.

\subsection{Hybrid tight-binding -- k$\cdot$p theory model of MLG and inelastic broadening of electron states}

The analysis of spectral form of SHG in this work is performed using the hybrid tight-binding -- $\vb{k}\cdot\vb{p}$ theory model Hamiltonian, where the intra-layer hopping of electrons across graphene lattices are accounted for by the Dirac model resulting from the $\vb{k}\cdot\vb{p}$ theory expansion around $\pm$K-points, whereas inter-layer couplings are included as hopping terms, in the spirit of tight-binding model that includes the full set of Slonczewski-Weiss-McClure parameters \cite{McClure_Band_1957,Slonczewsi_Band_1958}. With reference to the discussion of inversion symmetry breaking and $C_3$ trigonal symmetry of graphene above, it is natural to expect a crucial role played by terms in the MLG Hamiltonian that are responsible for trigonal warping effects in its dispersion. The hopping terms are represented graphically in Fig.~\ref{fig:tetralayer_bands} (their values are taken from Ref.~\cite{PhysRevB.104.085402}), and the analytical form of the Hamiltonian is given in SI Section 3. This MLG Hamiltonian is used to evaluate the SHG tensor component, $\sigma^a_{aa}(\omega)$, that -- according to Eq.~\eqref{eq:symmetry} -- fully sets the spectral form of SHG for all possible polarization choices. 

In addition to MLG bands dispersions and wave functions, the hybrid Hamiltonian provides information about the electron-photon coupling~\cite{ParkerMorimoto2019}, primarily, in the linear order in the photon field, as $\frac{e}{\hbar} A_a \frac{\partial H}{\partial k_a}$ and $\frac{e}{\hbar} A_a' \frac{\partial H}{\partial k_a}$, where $\vb{A}=\vb{E}^\omega/i\omega$ and $\vb{A'}=\vb{E}^{2\omega}/2i\omega$ are vector potentials of incoming and outgoing light, respectively. Here, we neglect higher-order electron-photon couplings~\cite{PhysRevB.108.085403}, $\frac{e^2}{\hbar^2} A_a^2 \frac{\partial^2 H}{\partial k_a^2}$, $\frac{e^2}{\hbar^2} A_aA_a' \frac{\partial^2 H}{\partial k_a^2}$ and $\frac{e^3}{\hbar^3} A_a^2 A_a' \frac{\partial^3 H}{\partial k_a^3}$, because those are much weaker as compared to the contributions coming from $v\pi$, $v_3\pi$ and $v_4\pi$ terms. This is why we represent the SHG process by a triangular Feynman diagram in Fig.~\ref{fig:diagramlatticepolarplot}(a), with single-photon vertices,
\begin{equation}
    e v^a \equiv  e\hbar^{-1}\partial H/\partial k_a.
\end{equation}
The triangular diagram in Fig.~\ref{fig:diagramlatticepolarplot}(a) highlights the fact that the SHG tensor is odd in the charge of the involved carrier. Therefore, SHG effect would be be absent in systems that feature an exact particle-hole symmetry. To mention, an approximate `minimal' MLG model, based solely on the closest-neighbor `vertical' $\gamma_1$-terms, has an artificial particle-hole symmetry~\cite{GarciaMcEllistrimFalko2023},
\begin{equation}
    U^{-1}H(\vb{k})U=-H(\vb{k}) , \quad U=1_N\otimes\sigma_z,
    \label{eq:phs}
\end{equation}
and its use would assume the absence of SHG even in the non-centrosymmetric MLG crystals, which is numerically demonstrated in SI Section 4 (here, $N$ is the number of layers, and the Pauli matrix $\sigma_z$ acts in the intra-layer A/B sublattice space). That is why the analysis of SHG characteristics requires taking into account next-neighbor McClure-Slonczewski-Weiss couplings known to lift the electron-hole symmetry in MLG, including $\gamma_2$, $\gamma_4$ and $\gamma_5$, as well as the on-site energy shift, $\Delta^\prime$, mutually induced on each other by the dimer sites orbitals.

Finally, all the above-listed structural block of the proposed theory are used to evaluate tensor component $\sigma^a_{aa}(\omega)$ using perturbation theory in the electron-photon coupling. Recently, such analyses of SHG were performed~\cite{ParkerMorimoto2019} using a generalized Kubo-formula approach extended onto non-linear optical responses. As a part of the formal derivations, such analyses involved analytical continuations of products of Green functions, in order to describe the SHG line shape in the near-resonant conditions, which would require special effort for appropriately taking into account intrinsic broadening of states in MLG bands (due to inelastic relaxation) rather than attributing broadening to the photons, or simply neglecting it ~\cite{BrunPedersen2015,Mikhailov2016,PassosVentura2018,PhysRevB.108.085403}. As the SHG is a kinetic process, we find it more natural to employ the Keldysh formalism for non-equilibrium Green's functions, which enables us to account appropriately for inelastic broadening, $\pm i\eta_n$, of all states, $\ket{n}$, involved in the SHG transitions. Note that while the closed-loop diagram in Fig.~\ref{fig:diagramlatticepolarplot}(a) indicates that the actual energy transfer in SHG occurs entirely in the photon field and graphene's charge carriers play only an intermediary role in the process (in contrast to electronic Raman scattering that leaves electron-hole pair behind~\cite{mcellistrim_spectroscopic_2023}), it also highlights the quantum nature of the SHG process, which is sensitive to inelastic broadening (decoherence) of electronic states due to kinetic processes intrinsic for graphene.  

An expression that follows from the Green-Keldysh functions analysis reads as, 

    \begin{align}
     \sigma^a_{aa}(\omega) = \frac{2e^3}{\omega^2}\int \frac{d^2k}{(2\pi)^2} \sum_{l,m,n}\frac{v^a_{ln}v^a_{nm}v^a_{ml}}{2\hbar\omega-\epsilon_{nl}+i(\eta_l+\eta_n)}\left(\frac{f(\epsilon_l) - f(\epsilon_m)}{\hbar\omega-\epsilon_{ml}+i(\eta_l + \eta_m)}-\frac{f(\epsilon_m) - f(\epsilon_n)}{\hbar\omega-\epsilon_{nm}+i(\eta_m + \eta_n)}\right),
    \label{eq:shgconducitivity}
\end{align}
where the full derivation is given in SI Section 5. Here, $\epsilon_{nm} = \epsilon_{n} - \epsilon_{m}$ are energies of transition between Bloch states $\ket{n, \vb{k}}$ and $\ket{m, \vb{k}}$, with energies $\epsilon_{n}$ and $\epsilon_{m}$, $f(\epsilon)$ is the Fermi function, and $v^a_{nm}$ are the matrix elements of the velocity operator in the $a$ direction between the involved states. As the higher-energy states (further away from the nominal monolayer Dirac point energy and the Fermi level in undoped MLG) experience stronger broadening due to inelastic electron-electron and electron-phonon scattering in graphene, in all numerical simulations below we assign broadenings $\eta_n \equiv \eta$ to all higher-energy bands shown in Fig.~\ref{fig:tetralayer_bands} and in a sketch on the l.h.s. of Fig.~\ref{fig:ABCB}, neglecting broadening (by setting $\eta_l \rightarrow 0$) for the states close to the conduction/valence band edge (e.g., bands $\pm1$ in ABCB tetralayer). As a result, spectra computed with Eq.~\eqref{eq:shgconducitivity} avoid spurious divergences at double resonances noticeable in the previous SHG theories~\cite{BrunPedersen2015,PhysRevB.108.085403}, which occur when the broadening is assigned to the photons rather than the electron states in the 2D material.

\section{Results}

\begin{figure*}[]
\centering
  { \includegraphics[width=1\textwidth]{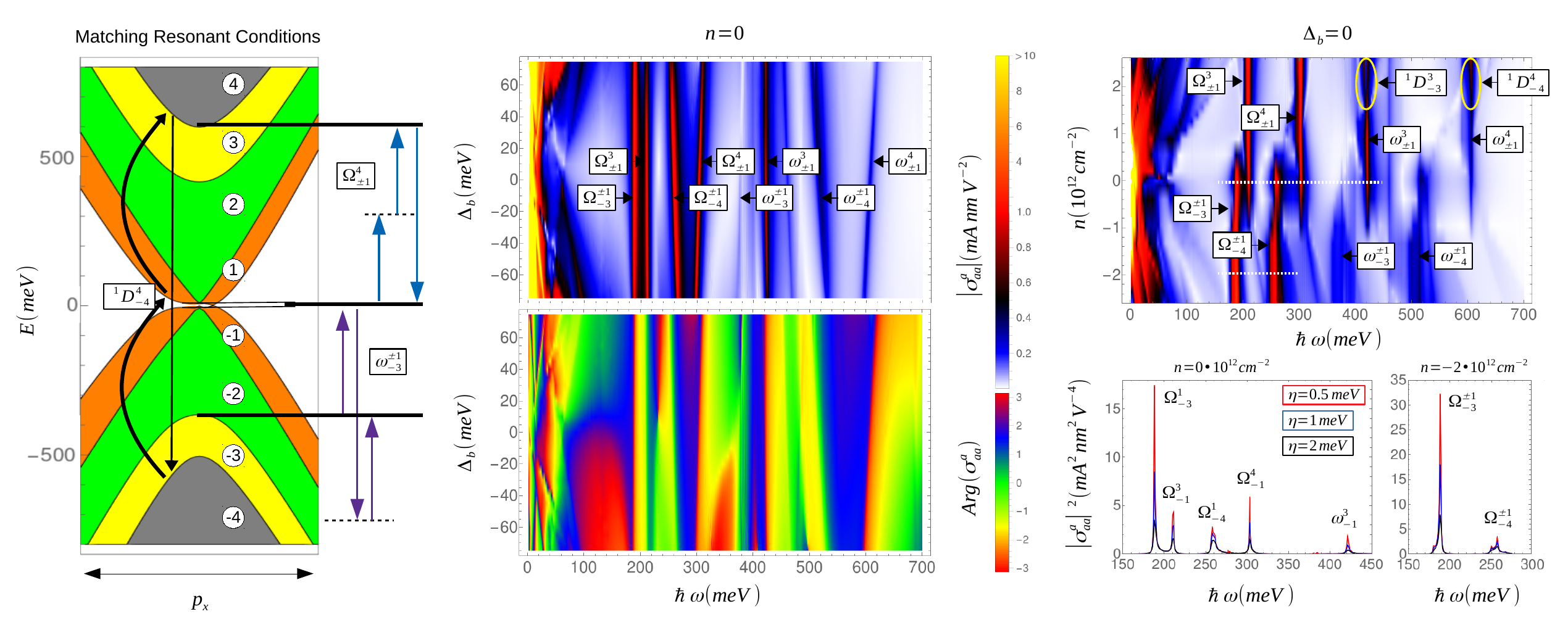}}
\caption{\label{fig:ABCB}
The color maps show the incoming photon frequency ($\omega$) dependence of the magnitude, $\abs{\sigma^a_{aa}}$ (top panels), and phase, ${\rm Arg}(\sigma^a_{aa})$ (bottom panels), of the SHG tensor in ABCB graphene. Vertical axes of those maps are used to show the variation of the SHG tensor with doping density, $n$ (middle column), and with a potential, $\Delta_b$, induced by the substrate on the bottom layer of MLG. On the left, we show an orthographic projection of the band structure of ABCB graphene (bands are enumerated using encircled indices), where we illustrate transitions involved in three types of resonant SHG. Here, solid horizontal lines point towards real states near the band edges involved in single-resonance transition, marked as $\omega^n_m$ and $\Omega^n_m$; thin dashed lines mark off-resonant virtual states. For double-resonant transitions, $^nD^m_l$, arrows point directly at the corresponding bands, identified by band indices $n,m,l$.}

\end{figure*}

\begin{figure*}[]
\centering
  { \includegraphics[width=1\textwidth]{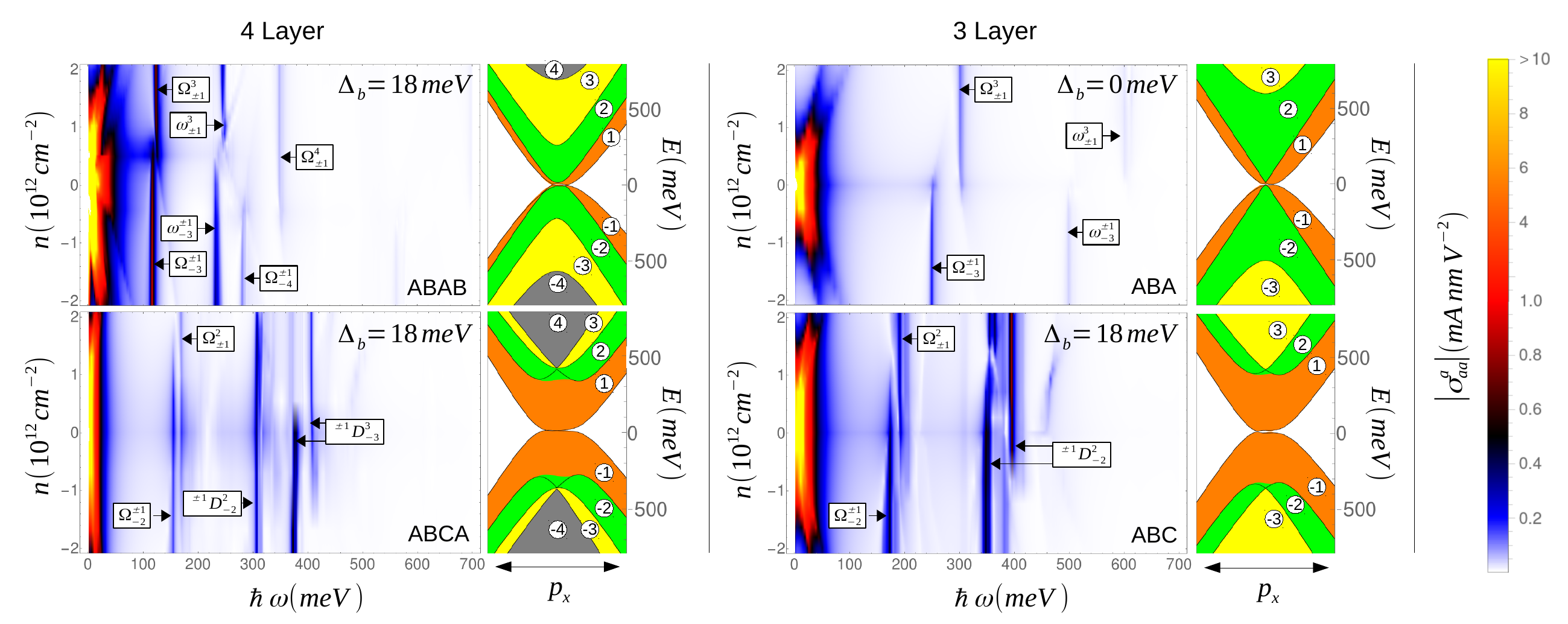}}
\caption{\label{fig:triandtetra}
We plot the intensity of SHG $\abs{\sigma^a_{aa}}$ for all remaining tetralayer and trilayer polytypes of graphene with stacking order labeled on the panels. Also labeled are the choice of $\Delta_b$, the externally induced inversion symmetry breaking parameter representing a substrate. We induce $\Delta_b=18{\rm meV}$ on the stacking orders that poses intrinsic inversion symmetry so as to induce a second order optical response. Along side each SHG intensity we plot the associated band structure with numbered bands. We label the high intensity features by resonance type and leading resonant band numbers using the same notation adopted in Fig.~\ref{fig:ABCB}. 
}
\end{figure*}

Below, we display the computed SHG spectral form, highlighting various resonant features of SHG in tetralayer and trilayer polytypes: primarily, in mixed stacking tetralayers (ABCB, in Fig.~\ref{fig:ABCB}) and ABA trilayers, where stacking order is the reason for inversion symmetry breaking, but also -- for comparison -- in asymmetrically perturbed Bernal and rhombohedral tetra- and trilayers (Fig.~\ref{fig:triandtetra}). The displayed spectra were all computed with Eq.~\eqref{eq:shgconducitivity} using the Hamiltonians shown in SI Section 3 and illustrated in Fig.~\ref{fig:tetralayer_bands} in terms of the included inter-layer hopping. We also pay attention to the influence of an encapsulation environment (via on-layer potential shifts in the bottom layer, $\Delta_{b}$) and doping density, $n$. The resonances in the computed spectra reflect the MLG electronic band structures shown in Fig.~\ref{fig:tetralayer_bands} and by orthographic projections in Figs.~\ref{fig:ABCB} and \ref{fig:triandtetra}, with $\eta \approx 1$meV.

The color maps presented in Fig.~\ref{fig:ABCB} highlight the resonant structure of SHG in the mixed-stacking (ABCB) tetralayer. The two panels on the top, that show the SHG amplitude, feature pronounced peaks, accompanied by $\pi$-jumps of the SHG phase (bottom panel in the middle column). Each of the identified resonances in the top panels is attributed a particular sequence of transitions between the bands, marked on the orthographic projection on the l.h.s. of Fig.~\ref{fig:ABCB}. Here, we identify two types of features: those corresponding to pairs of bands in resonance with single photon energy (labeled as $\omega^l_n$), and those resonant with a $2\omega$ transition (labeled as $\Omega^l_n$). For a better understanding of those resonances, we rewrite Eq.~\eqref{eq:shgconducitivity}, separating the terms responsible for the corresponding peaks:

    \begin{align}
    \sigma^a_{aa}(\omega) &= \frac{2e^3}{\omega^2}\int \frac{d^2k}{(2\pi)^2} \sum_{l,m,n}v^a_{ln}v^a_{nm}v^a_{ml}\left[\frac{1}{\epsilon_{ml}-\epsilon_{nm}-i\delta_{lmn}}\frac{f(\epsilon_l)-f(\epsilon_m)}{\hbar\omega-\epsilon_{ml}+i(\eta_l+\eta_m)}\right. \nonumber
    \\
    &\hspace{5.cm}+\frac{1}{\epsilon_{ml}-\epsilon_{nm}+i\tilde{\delta}_{lmn}}\frac{f(\epsilon_m)-f(\epsilon_n)}{\hbar\omega-\epsilon_{nm}+i(\eta_n+\eta_m)} \nonumber
    \\
    &\hspace{5.cm}\left.-2\left(\frac{f(\epsilon_l)-f(\epsilon_m)}{\epsilon_{ml}-\epsilon_{nm}-i\delta_{lmn}}+\frac{f(\epsilon_m)-f(\epsilon_n)}{\epsilon_{ml}-\epsilon_{nm}+i\tilde{\delta}_{lmn}}\right)\frac{1}{2\hbar\omega-\epsilon_{nl}+i(\eta_l+\eta_n)}\right].
    \label{eq:shgconducitivity2}
\end{align}

Here, $\delta_{lmn}~=~\eta_l+2\eta_m-\eta_n$ and $\tilde{\delta}_{lmn}~=~-\eta_l+2\eta_m+\eta_n$. The first and second rows in Eq.~\eqref{eq:shgconducitivity2} capture the $\omega^n_l$-type resonances, whereas the third row highlights $\Omega^n_l$-type resonances. 

Note that these SHG resonances are determined by the coherent convolution of inelastically broadened transitions with the optical density of states across the continuum spectrum of electronic bands, so that the resulting lineshapes are not Lorentzian. This point is demonstrated by the comparison of the SHG lineshapes computed for several values of $\eta$, see bottom right panel in Fig.~\ref{fig:ABCB}, indicating that the reduction of single-particle broadening does not lead to the overall narrowing of the SHG peak: instead, it develops a sharp asymmetric spike originating from van Hove singularities in the bands $\pm1$ near the conduction/valence band edge of the tetralayer. To highlight those spectral characteristics of each type of resonance, we develop an approximate analytical description of their intensity. For this, we approximate inter-band splittings in the dominant order of $\gamma_1$, and account the main relevant contributions towards the velocity matrix elements between ABCB MLG bands at $\vb{k}=0$, parametrized by $v$ and $v_3$, in Eq. (9). 

\begin{equation} \label{eq:ABCB}
||\bra{m}\hat{v}^a (\vb{k}\rightarrow 0)\ket{n} ||=
\begin{pmatrix}

 0&\frac{-i\sqrt{2}v}{4} & \frac{-iv}{2} & \frac{-iv_3}{2} & \frac{-i\sqrt{2}v}{2} & 0 & \frac{-i\sqrt{2}v}{4} & 0\\
\frac{i\sqrt{2}v}{4} & 0 & 0 & \frac{i\sqrt{2}v}{2} & \frac{i\sqrt{2}v_3}{2} & \frac{-iv}{2} & 0 & \frac{i\sqrt{2}v}{4}\\
\frac{iv}{2} & 0 & 0 & 0 & i v_3 & \frac{i\sqrt{2}v}{2} & 0 & \frac{iv}{2} \\
\frac{iv_3}{2} & \frac{-i\sqrt{2}v}{2} & 0 & 0 & 0 & \frac{-i\sqrt{2}v_3}{2} & \frac{i\sqrt{2}v}{2} & \frac{iv_3}{2}\\
\frac{i\sqrt{2}v}{2} & \frac{-i\sqrt{2}v_3}{2} & -iv_3 & 0 & 0 & 0 & \frac{-i\sqrt{2}v_3}{2} & \frac{-i\sqrt{2}v}{2}\\
0 & \frac{iv}{2} & \frac{-i\sqrt{2}v}{2} & \frac{i\sqrt{2}v_3}{2} & 0 & 0 & \frac{iv}{2} & 0\\
\frac{i\sqrt{2}v}{4} & 0 & 0 & \frac{-i\sqrt{2}v}{2} & \frac{i\sqrt{2}v_3}{2} & \frac{-iv}{2} & 0 & \frac{-i\sqrt{2}v}{4}\\
0 & \frac{-i\sqrt{2}v}{4} & \frac{-iv}{2} & \frac{-iv_3}{2} & \frac{i\sqrt{2}v}{2} & 0 & \frac{i\sqrt{2}v}{4} & 0
\end{pmatrix}.
\end{equation}

For the $\Omega$-type resonances identified in Fig.~\ref{fig:ABCB} at zero charge doping, $\Omega_{-3}^{\pm1},\Omega_{\pm1}^{3}$ and $\Omega_{-4}^{\pm1},\Omega_{\pm1}^{4}$, we find:
\begin{equation}
 \frac{I^{2\omega}_{\rm out}}{\abs{I^\omega_{\rm in}}^2} \approx \alpha_{n}^l\frac{A\hbar^4 e^6v^4 v_3^2}{\gamma_1^6(\varepsilon_0c)^3}\abs{\int \frac{d^2k}{(2\pi)^2}\frac{1}{2\hbar\omega-\Omega_n^l+i\eta}}^2,
 \nonumber
\end{equation}
with numerical factors, $\alpha_{n}^l$, listed in Table I.  Note that both $v$ and $v_3$ elements of the velocity matrix enter the above expression, reflecting the role of trigonal warping in the MLG lattice structure. 
These $\Omega$-type features are robust against changes in the encapsulation environment $\Delta_b$, as shown in the middle column of Fig.~\ref{fig:ABCB}.
In contrast, as we show in the top right panel of Fig.~\ref{fig:ABCB}, these resonant spectra are significantly influenced by doping, $n$, which is determined by Pauli blocking of transitions to/from bands close to the conduction/valence band edge.

\begin{table}
\begin{center}
\begin{tabular}{|c|c|c|c|c|c|c|c|c|c|c|}
\hline
Res &
$\Omega_{-3}^{1}$& $\Omega_{-1}^{3}$& $\Omega_{-4}^{1}$& $\Omega_{-1}^{4}$&
$\omega_{-3}^{1}$& $\omega_{-1}^{3}$& $\omega_{-4}^{1}$& $\omega_{-1}^{4}$&
$^1D_{-3}^{3}$& $^1D_{-4}^{4}$\\  
\hline
 $\alpha_n^l $& $447$& $126$& $424$& $216$&
$3$& $42$& $34$& $6$& $190$ & $306$\\
\hline
\end{tabular}
\caption{The numerical factor $\alpha_n^l$ used to describe the simplified forms of all three types of resonance. These are specific for all relevant band transitions identified in Fig.~\ref{fig:ABCB} relating to ABCB graphene with zero doping.}
\end{center}
\end{table}

For the $\omega$-type resonant features, the simplified formula for the SHG lineshape is:
\begin{equation}
 \frac{I^{2\omega}_{\rm out}}{\abs{I^\omega_{\rm in}}^2} \approx \alpha_{n}^l\frac{A\hbar^4 e^6v^4 v_3^2}{\gamma_1^6(\varepsilon_0c)^3}\abs{\int \frac{d^2k}{(2\pi)^2}\frac{1}{\hbar\omega-\omega_n^l+i\eta}}^2,
 \nonumber
\end{equation}
with numerical factors $\alpha_{n}^l$ listed in Table I. As for $\Omega$ resonances, both $v$ and $v_3$ elements of the velocity matrix enter the above expression, reflecting the role of trigonal warping in the MLG lattice structure, and the $\omega$-resonances are affected by doping, $n$, due to Pauli blocking of transitions to/from bands close to the conduction/valence band edge.

To mention, there are also two `double-resonance' features $^{1}D^{3}_{-3}$ and $^{1}D^{4}_{-4}$ in Fig.~\ref{fig:ABCB}, attributed to the simultaneous resonance of both $\omega$ and $2\omega$ transitions, which happens when the process involves three almost equally spaced band edges ~\cite{RosencherBoisNagle1996,FrigerioVirgilioOrtolani2021}. In Fig.~\ref{fig:ABCB}, we notice only two such resonances, which appear only at doping $n \ge 1.5\times10^{12}\rm{cm}^{-2}$, which is because states shifted slightly away from the band edges are needed to satisfy the energy-matching condition in the ABCB tetralayer. To describe $^{1}D^{3}_{-3}$ and $^{1}D^{4}_{-4}$ resonance analytically (details given in SI Section 5, also see in Ref.~\cite{RosencherBoisNagle1996,FrigerioVirgilioOrtolani2021}), we simplify Eq.~\eqref{eq:shgconducitivity2} into:
\begin{equation}
 \frac{I^{2\omega}_{\rm out}}{\abs{I^\omega_{\rm in}}^2} \approx \alpha_{n}^l\frac{A\hbar^4 e^6v^4 v_3^2}{\gamma_1^4(\varepsilon_0c)^3}\abs{\int \frac{d^2k}{(2\pi)^2}\frac{1}{(\hbar\omega-^{m}D^{l}_{n}+i\eta)^2}}^2.
 \nonumber
\end{equation}
Markedly, the form of the pole in the integrand of the above equation differs from what appears in the simplified expressions for the single resonance features, enhancing its dependence of the broadening of electronic states and sharpening up the resonance for smaller $\eta$'s.

Similar resonance features can also be noticed in the computed spectra of other tetralayers and, also, the trilayers in Fig.~\ref{fig:triandtetra}. Next to each color map, we show the corresponding orthographic projection of the band structure and identify band indices. Note that, while for ABA trilayer, which lattice structure in non-centrosymmetric, the SHG effect is intrinsic, for the modeled ABC trilayers and two centrosymmetric tetralayers, the inversion symmetry breaking is caused by the substrate via a proximity potential, $\Delta_b$. Here, we  choose $\Delta_b=18$meV, with the reference to MLG on a misaligned hBN substrate~\cite{boschi_built-bernal_2024}. As in the ABCB tetralayer, each of other polytypes exhibits a series of $\Omega$-type and $\omega$-type resonances in the infra-red regime. We also note that rhombohedral multilayers show very clear double-resonant peaks (absent in Bernal MLGs), which we attribute to the pronounced van Hove singularities around the neutrality point, which is also a feature of the ABCB tetralayer spectrum.

\section{Discussion and Conclusions}

In summary, we calculated the second harmonic generation (SHG) in five different multilayer graphene (MLG) polytypes using a Green-Keldysh formulation, identifying resonant spectral feature that can be used for optical identification of various MLG polytypes.
The analysis was performed from the terahertz (THz) to the infra-red (IR) regime. While SHG in the THz range seems to be strong in ABCB tetralayer and ABA trilayer, it is generally featureless and also appears in other polytypes if inversion symmetry of those is lifted by the encapsulation environment; it is also sensitive to the MLG doping, hence, making it difficult to use as a probe of the MLG stacking configuration. In contrast, the clearly identifiable peaks in the IR range of SHG, classified as the resonances $\omega_n^m$ related to the input frequency ($\omega$), the resonances $\Omega_n^m$ at the output frequency ($2\omega$), and double-resonances, $^lD_n^m$ (a simultaneous activation of the former two) can provide a unique identification of the MLG polytypes.

\section*{Acknowledgements}

PJS acknowledges support from CDT Graphene-NOWNANO, and VF from the British Council - ISPF grant 1185409051.
TVJ, TK and MK were supported by JSPS KAKENHI Grants No. JP20K14415, No. JP20H01840, No. JP20H00127, No. JP21H05236, No. JP21H05232, JP24K06921, by JST CREST Grant No. JPMJCR20T3, and by JST SPRING, Grant No. JPMJSP2138. 
The collaboration between Manchester and Osaka was supported by NPL - ISPF project QEMT.

\printbibliography

\end{document}